\documentclass[11pt]{article} 
\newcommand{\be}{\begin{equation}}
\newcommand{\ee}{\end{equation}}
\newcommand{\bea}{\begin{eqnarray}}
\newcommand{\eea}{\end{eqnarray}}
\newcommand{\sn}{{\rm sn}}
\newcommand{\ds}{{\rm ds}}
\newcommand{\cs}{{\rm cs}}
\newcommand{\ns}{{\rm ns}}
\newcommand{\dn}{{\rm dn}}
\newcommand{\cn}{{\rm cn}}
\newcommand{\sech}{{\rm sech}}

\begin{document}
\vspace{.5in} 
\begin{center} 
{\LARGE{\bf Periodic and Hyperbolic Soliton Solutions of a Number of 
Nonlocal PT-Symmetric  Nonlinear Equations}}
\end{center} 

\vspace{.3in}
\begin{center} 
{\LARGE{\bf Avinash Khare}} \\ 
{Raja Ramanna Fellow,}
{Indian Institute of Science Education and Research (IISER), Pune, India
411021}
\end{center} 

\begin{center} 
{\LARGE{\bf Avadh Saxena}} \\ 
{Theoretical Division and Center for Nonlinear Studies, Los
Alamos National Laboratory, Los Alamos, NM 87545, USA}
\end{center} 

\vspace{.9in}
{\bf {Abstract:}}  

For a number of nonlocal nonlinear equations such as nonlocal, nonlinear 
Schr\"odinger equation (NLSE), nonlocal Ablowitz-Ladik (AL), nonlocal, 
saturable discrete NLSE (DNLSE), coupled nonlocal NLSE,  coupled nonlocal 
AL and coupled nonlocal, saturable DNLSE, we obtain periodic solutions 
in terms of Jacobi elliptic functions as well as the corresponding hyperbolic
soliton solutions. Remarkably, in all the six cases, we find that unlike the 
corresponding local cases, all the nonlocal models simultaneously admit both 
the bright and the dark soliton solutions. Further, in all the six cases, not 
only $\dn(x,m)$ and $\cn(x,m)$ but even their linear superposition is shown to be an 
exact solution. Finally, we show that the coupled nonlocal NLSE not only 
admits solutions in terms of Lam\'e polynomials of order 1, but it also admits 
solutions in terms of Lam\'e polynomials of order 2, even though they are not 
the solutions of the uncoupled nonlocal problem.  We also remark on the 
possible integrability in certain cases. 

\newpage 
  
\section{Introduction} 

In recent years non-Hermitian but PT-symmetric systems have attracted widespread 
attention \cite{ben}. In the context of the nonrelativistic quantum mechanics, it has 
been shown that these systems exhibit real spectra so long as the PT-symmetry is not
spontaneously broken. On the other hand if PT-symmetry is spontaneously broken, then
the spectrum is not entirely real. In the last few years, it has been realized 
that optics can
provide an ideal ground for testing some of the consequences of such theories.
This is because, the paraxial equation of diffraction is similar in structure
to the Schr\"odinger equation. Several studies have in fact shown that 
PT-symmetric optics could give rise to an entirely new class of optical
structures and devices with altogether new properties \cite{avadh}.  It is thus 
imperative to study different types of PT-invariant nonlinear systems which
could have implications in optics. 

In an interesting recent paper, Ablowitz and Musslimani \cite{am} 
(here after, we refer to it as I) have 
considered nonlocal nonlinear Schr\"odinger equation (NLSE) which is 
non-Hermitian but PT-invariant and have shown that it is an integrable system.
Further, Musslimani and his collaborators \cite{msmc} (here after we refer
to it as II) have shown that unlike
the usual NLSE, the nonlocal NLSE remarkably admits both bright and 
dark soliton solutions. In addition, they have also noted that unlike the local
case, these solitons do not admit solutions with arbitrary shift in the
transverse coordinate $x$. 

The purpose of this paper is to study a number of other nonlocal
nonlinear equations, all of which are non-Hermitian but PT-invariant and enquire
if they too simultaneously admit both bright and dark soliton solutions. In 
particular, we study nonlocal Ablowitz-Ladik (AL)  model, nonlocal coupled
AL model, nonlocal saturable discrete NLSE (DNLSE), nonlocal coupled
saturable DNLSE  as well as nonlocal coupled NLSE models and obtain 
periodic as well as hyperbolic soliton solutions in all these cases. 
We show that unlike the corresponding local models, all these 
nonlocal models, for the focusing type of nonlinearity, 
simultaneously admit both the bright and the dark soliton solutions. As far 
as we are aware 
of, this is the first time that one has models simultaneously admitting both
the dark and the bright soliton solutions. We also show that while these 
nonlocal models do not admit solutions with arbitrary shift in the transverse
coordinate $x$ \cite{msmc}, they do admit solutions with specific shifts. 
Besides, all these models not only admit periodic solutions in terms of Jacobi 
elliptic functions $\dn(x,m)$  and $\cn(x,m)$ with modulus $m$, but even their 
linear superposition \cite{ks} is also an exact periodic 
solution in all these cases. Further, in the coupled nonlocal NLSE models, we show 
that they not only admit periodic and hyperbolic soliton solutions in terms of 
Lam\'e polynomials of order 1, 
but they also admit periodic and hyperbolic soliton solutions in terms of Lam\'e 
polynomials of order 2, even though they are not the solutions of the uncoupled 
problem. Besides, the coupled model also admits periodic solutions in terms of 
shifted
Lam\'e polynomials of order 1 and 2. We also obtain similar periodic soliton 
solutions in the case of the various discrete models, i.e. nonlocal AL,
coupled nonlocal AL, nonlocal saturable DNLSE and coupled nonlocal  
saturable DNLSE models.

The plan of the paper is as follows. In Sec. II we consider nonlocal NLSE 
introduced in I and II, and obtain period solutions in terms of not only $\cn(x,m)$ 
and $\dn(x,m)$ but also their linear superposition. Further, in the same focusing 
model we also obtain periodic solutions in terms of $\sn(x,m)$ and hence the dark
and the bright hyperbolic soliton solutions. 
Finally, we show that this model also admits solutions in terms of
shifted $\sn(x,m)$, $\cn(x,m)$ and $\dn(x,m)$. In Sec. III we consider coupled nonlocal NLSE
model and obtain periodic solutions in terms of Lam\'e polynomials of order 1 and 2. We show 
that while both nonlocal Manakov and nonlocal Mikhailov-Zakharov-Schulman (MZS) models 
admit periodic solutions in terms of Lam\'e polynomials of order 1, only MZS case admits 
periodic solutions in terms of Lam\'e polynomials of order 2 while in the Manakov limit,
there are no solutions in terms of Lam\'e polynomials of order 2. However, we
also show that the shifted Lam\'e polynomials of order 1 and 2 do admit 
solutions in both the Manakov and the MZS cases.  
In Sec. IV we consider a nonlocal saturable DNLSE model and show that the same 
focusing model simultaneously admits both $\cn(x,m)$ and $\dn(x,m)$ as well as $\sn(x,m)$ 
solutions and hence the corresponding hyperbolic bright and the dark soliton solutions. 
Further, we show that the same model, also admits shifted 
$\sn(x,m)$, $\cn(x,m)$ and $\dn(x,m)$ solutions. 
In Sec. V we consider a coupled nonlocal saturable DNLSE and obtain
solutions in terms of Lam\'e polynomials of order 1. In Sec. VI we study the 
nonlocal AL model and show that even this model admits both the dark and the 
bright as well as the superposed periodic solutions.  Finally, this model too admits 
shifted $\sn(x,m)$, $\cn(x,m)$ 
and $\dn(x,m)$ solutions. In Sec. VII we consider the coupled nonlocal AL model and 
obtain
periodic solutions in terms of Lam\'e polynomials of order 1. In the last section we 
summarize our main results and speculate about the integrability of the 
nonlocal AL, nonlocal Manakov and nonlocal MZS models.

\section{Periodic Solutions of Nonlocal NLSE}

We start with the nonlocal NLSE as considered in I and II
\be\label{1}
iu_{z}(x,z) +u_{xx}(x,z)+ g u(x,z) u^{*}(-x,z)u(x,z)  =0\,.
\ee
As has been remarked in II, one can look upon the nonlinear term as a 
self-induced potential of the form $V(x,z) = u(x,z)u^{*}(-x,z)$ which is
PT-invariant in the sense that $V(x,z) = V^{*}(-x,z)$. As shown in II,
this is a non-Hermitian system where total power defined by
\be\label{2}
P = \int_{-\infty}^{\infty} dx\, |u(x,z)|^2\,,
\ee
is not conserved but it is the quasi-power $Q$ and the Hamiltonian $H$ given by
\be\label{3}
Q = \int_{-\infty}^{\infty} dx\, u(x,z)u^{*}(-x,z)\,,~~
H= \int_{-\infty}^{\infty} dx\,[u_{x}(x,z)u^{*}_{x}(-x,z)
-\frac{1}{2} u^2(x,z){u^{*}}^{2}(-x,z)]\,,
\ee
which are conserved.

Before obtaining the periodic solution, notice that the nonlocal NLS Eq. (\ref{1}) 
has a novel (plane wave) exact solution
\be\label{4a}
u(x,z) = A\,e^{kx -i(\omega z+\delta)}\,,
\ee
provided
\be\label{4b}
g A^2 = -(\omega+k^2)\,.
\ee
Thus solution (\ref{4a}) holds good irrespective of whether $g > 0$ or $g < 0$. 

Let us now obtain periodic soliton solutions to this equation. 

{\bf Solution I}

It is easily checked that one of the exact periodic solution to this equation is
\be\label{4}
u = A \dn(\beta x,m) e^{-i(\omega z+\delta)}\,,
\ee
provided
\be\label{5}
g A^2 = 2 \beta^2\,,~~\omega = -(2-m)\beta^2\,.
\ee
Here $\delta$ is an arbitrary constant. 
Thus for this solution $-2 < \omega \le -1$ as $0 < m \le 1$, 

{\bf Solution II}

Similarly, another exact periodic solution to the nonlocal NLS Eq. (\ref{1}) is
\be\label{6}
u = A \sqrt{m}\, \cn(\beta x,m) e^{-i(\omega z+\delta)}\,,
\ee
provided
\be\label{7}
g A^2 = 2 \beta^2\,,~~\omega = -(2m-1)\beta^2\,.
\ee
Thus for this solution $-1 \le \omega < 1$ as $0 < m \le 1$, 

{\bf Solution III}

Remarkably, even a linear superposition of the two, i.e.
\be\label{8}
u = \left [ \frac{A}{2} \dn(\beta x,m)
+ \frac{B}{2} \sqrt{m}\, \cn(\beta x,m) \right ]e^{-i(\omega z+\delta)}\,,
\ee
is also an exact periodic solution to the nonlocal NLS Eq. (\ref{1}) provided
\be\label{9}
B = \pm A\,,~~gA^2 = 2\beta^2\,,~~\omega = -(1/2)(1+m)\beta^2\,.
\ee
Thus for this solution $-1 \le \omega < -1/2$ as $0 < m \le 1$, 

{\bf Solution IV}

In the limit $m=1$, all three periodic solutions, i.e. $\cn,\dn$ as well
as superposed solution with $B=+A$, go over to the same hyperbolic bright 
soliton solution 
\be\label{10}
u = A \sech(\beta x) e^{-i(\omega z+\delta)}\,,
\ee
provided
\be\label{11}
g A^2 = 2 \beta^2\,,~~\omega = -\beta^2\,.
\ee
On the other hand, the superposed solution (\ref{6}) with $B=-A$ goes over to 
the vacuum solution $u=0$.

{\bf Solution V}

Remarkably, unlike the NLSE case, the same (focusing) nonlocal NLS Eq. (\ref{1}) 
also admits the periodic $\sn$ solution
\be\label{12}
u = A \sqrt{m}\, \sn(\beta x,m) e^{-i(\omega z+\delta)}\,,
\ee
provided
\be\label{13}
g A^2 = 2 \beta^2\,,~~\omega = (1+m)\beta^2\,.
\ee
Thus for this solution $1 < \omega \le 2$ as $0 < m \le 1$, 

At this point, it is worth asking the following question: just as there 
are periodic solutions
$\cn$ and $\dn$ with period $4K(m)$ and $2K(m)$ respectively both of which go 
over to the same pulse solution ($\sech$) in the $m=1$ limit, why one only has 
$\sn$ solution with period $4K(m)$ but no period $2K(m)$ solution with both 
going to the same dark soliton solution in the  $m=1$ limit? Such a solution 
is in fact possible. In particular, it is easily checked that
\be\label{12a}
u = A m\, \frac{\cn(\beta x,m) \sn(\beta x,m)}{\dn(\beta, x)}\, 
e^{-i(\omega z+\delta)}\,,
\ee
is an exact solution (with period $2K(m)$) to Eq. (\ref{1}) provided
\be\label{13a}
g A^2 = 2 \beta^2\,,~~\omega = 2(2-m)\beta^2\,.
\ee
Thus for this solution $2 \le \omega < 4$ as $0 < m \le 1$, 
Note that at $m=1$, both solutions (\ref{12}) and (\ref{12a}) go over to the same 
kink solution. Perhaps the reason why this solution is not considered in the 
literature is because it is related to the $\sn$ solution by Landen transformation 
\be\label{12b}
(1+\sqrt{1-m})\frac{\sn(x,m)\cn(x,m)}{\dn(x,m)} =
\sn\left[(1+\sqrt{1-m})x,\left(\frac{1-\sqrt{1-m}}{1+\sqrt{1-m}}\right)^{2}\right]\,.
\ee

{\bf Solution VI}

In the limit $m=1$, the periodic solution V (and the one given by Eq. 
(\ref{12a})  go over to the same hyperbolic dark soliton solution
\be\label{14}
u = A \tanh(\beta x) e^{-i(\omega t+\delta)}\,,
\ee
provided
\be\label{15}
g A^2 = 2 \beta^2\,,~~\omega =  2\beta^2\,.
\ee

It is worth pointing out that while the defocusing NLSE ($g< 0$) 
admits $\sn$ solution, defocusing, nonlocal NLSE (i.e. Eq. (\ref{1}) 
with $g<0$) does not
admit a periodic solution in terms of either $\sn$ or $\sn \cn/dn$(
or $\cn$ or $\dn$) or the 
hyperbolic kink or pulse solutions. However, as we now show, in the defocusing
case, there are solutions in terms of shifted $\sn$ and $\cn$.

\subsection{Shifted $\sn,\cn,\dn$ Solutions}

If we look at all the periodic as well as the hyperbolic soliton solutions 
of the nonlocal equations, we find that unlike the local NLSE case, the 
solutions of the nonlocal NLSE are {\it not invariant} with respect to 
shifts in the transverse coordinate $x$. For example, while 
$A \dn[\beta(x+x_{0}),m]e^{-i\omega t}$ is an exact solution of the local NLSE,
no matter what $x_{0}$ is, as remarked in II, for arbitrary $x_{0}$,
it is not an exact solution of our nonlocal NLSE. However, we now show that
for special values of $x_{0}$, $\sn,\cn,\dn$ are still the solutions of the
nonlocal NLSE. In particular, we show that when $x_{0} = K(m)$, there are
exact solutions of the nonlocal NLSE as given by Eq. (\ref{1}) for both
the focusing ($g > 0$) and the defocusing ($g < 0$) cases. This is because,
\be\label{16h}
\dn[x+K(m),m]=\frac{\sqrt{1-m}}{\dn(x,m)}\,,~~ 
\sn[x+K(m),m]=\frac{\cn(x,m)}{\dn(x,m)}\,,~~ 
\cn[x+K(m),m]= - \sqrt{1-m} \frac{\sn(x,m)}{\dn(x,m)}\,,
\ee
and as we show now, these are indeed the solutions of the nonlocal NLSE
Eq. (\ref{1}). 

{\bf Solution VII}

It is easily checked that an exact periodic solution to the nonlocal
NLSE (\ref{1}) is
\be\label{2.1}
u = A \frac{\sqrt{1-m}}{\dn(\beta x,m)} e^{-i(\omega z+\delta)}\,,
\ee
provided conditions as given by Eq. (\ref{5}) are satisfied, which are 
precisely the conditions under which $\dn$ is an exact solution.

{\bf Solution VIII}

Yet another periodic solution to Eq. (\ref{1}) is 
\be\label{2.2}
u = A \sqrt{m(1-m)} \frac{\sn(\beta x,m)}{\dn(\beta x,m)} 
e^{-i(\omega z+\delta)}\,,
\ee
provided
\be\label{2.3}
gA^2 = -2\beta^2\,,~~\omega = -(2m-1) \beta^2\, . 
\ee

{\bf Solution 	IX}

Yet another exact periodic solution to Eq. (\ref{1}) is 
\be\label{2.4}
u = A \sqrt{m} \frac{\cn(\beta x,m)}{\dn(\beta x,m)} 
e^{-i(\omega z+\delta)}\,,
\ee
provided
\be\label{2.5}
(1-m) gA^2 = -2(1-m) \beta^2\,,~~\omega = (1+m) \beta^2\, . 
\ee

Note that while $\cn$ and $\sn$ are the exact solution in the focusing case, 
i.e. $g > 0$, the above two solutions (\ref{2.2}) and (\ref{2.4}) (which are 
just the shifted $\cn$ and $\sn$ solutions respectively), 
are however only valid in the defocusing case, i.e.  $g < 0$. We find this to 
be rather interesting. Further, while $\dn, \cn$ and
$\sn$ are the exact solutions of our nonlocal NLSE for all nonzero values of 
$m$ including $m=1$ (i.e. for hyperbolic solitons), the above three 
shifted periodic solutions are only valid in case $0 < m < 1$. 

It may be noted here that there are solutions to the nonlocal NLSE Eq. 
(\ref{1}) even when $x_{0} = iK(1-m)$ or when $x_{0} = K(m)+iK(1-m)$ but
these are singular solutions and hence we do not consider them in this
paper.

Summarizing, we find that for the nonlocal NLSE with focusing type ($g > 0$) 
of nonlinearity, while 
$\cn,\dn,\sn$ and $1/\dn$ are the exact periodic solutions, for the defocusing 
case (i.e. $g<0$), $\cn/\dn$ and $\sn/\dn$ are the exact periodic solutions.

\section{Nonlocal Coupled NLSE}

Let us consider the following coupled nonlocal NLS field equations
\bea\label{3.1}
&&iu_{z}(x,z)+u_{xx}(x,z)+[a u(x,z) u^{*}(-x,z) +b v(x,z)v^{*}(-x,z)] u(x,z) =0 , 
\nonumber \\
&&iv_{z}(x,z)+v_{xx}(x,z)+[f u(x,z) u^{*}(-x,z) +e v(x,z)v^{*}(-x,z)] v(x,z) =0 , 
\eea
where $u$ and $v$ are the two coupled nonlocal NLS fields and $a,b,f,e$ are 
arbitrary real numbers. 
This is also a non-Hermitian PT-invariant system in the sense that the self-induced 
potential $V(x,z) = u(x,z)u^{*}(-x,z)+v(x,z)v^{*}(-x,z)$ satisfies 
$V(x,z) = V^{*}(-x,z)$. In this model, whereas total power, defined by
\be\label{3.1a}
P = \int_{-\infty}^{\infty} dx\, (|u(x,z)|^2+|v(x,z)|^2)\,,
\ee
is not conserved, but the quasi-powers given by
\bea\label{3.1b}
&&Q_{1} = \int_{-\infty}^{\infty} dx\, [u(x,z)u^{*}(-x,z)]\,,
\nonumber \\
&&Q_{2} = \int_{-\infty}^{\infty} dx\, [v(x,z)v^{*}(-x,z)]\,,
\eea
are separately conserved. In the special case when $a=f,b=e$, the so called mixed
quasi-power given by
\be\label{3.1c}
Q_{3} = \int_{-\infty}^{\infty} dx\, [u(x,z)v^{*}(-x,z)+v(x,z)u^{*}(-x,z)]\,,
\ee
is also conserved. Further, 
there is another conserved quantity 
\bea\label{3.1d}
&&C_{1} = u_{x}(x,z)u^{*}_{x}(-x,z)+v_{x}(x,z)v^{*}_{x}(-x,z) \nonumber \\
&&-\frac{1}{2}[a u^{2}(x,z)(u^{*}(-x,z))^{2} +e v^{2} (x,z)(v^{*}(-x,z))^{2} 
- 2g u(x,z)u^{*}(-x,z)v(x,z)v^{*}(-x,z)]\,.
\eea
provided $g=b=f$. In particular, in the so called Manakov limit 
(i.e. $a=b=f=e$), we have the conserved quantity $C_{1}$. 

It is worth pointing out that these conserved  quantities are similar to the 
corresponding conserved quantities in the local coupled NLSE, in particular,
the conserved quantities for the nonlocal case are obtained from the corresponding
local model by simply replacing $u^{*}(x,z)$ and $v^{*}(x,z)$ by $u^{*}(-x,z)$ and
$v^{*}(-x,z)$, respectively. 

We now show that these coupled equations admit periodic solutions in terms of Lam\'e
polynomials of order 1  and 2. In the special case
when $a=f=b=e$ this system reduces to the nonlocal Manakov system. In this context, 
it is worth remembering that 
the corresponding local Manakov system is a well known integrable system 
\cite{man}. Remarkably, even when $a=f=-b=-e$, the local case corresponds to the 
integrable MZS system \cite{mik,zak,ger}. Hence we shall call the above coupled
equations as nonlocal MZS system in case $a=f=-b=-e$. 
We shall however discuss the exact periodic solutions of this coupled nonlinear 
system when the coefficients $a,b,f,e$ are arbitrary but real. 

Before we discuss the periodic solutions to the coupled Eqs. (\ref{3.1}), we
remark that the coupled system has a novel exact solution given by
\be\label{3.1g}
u(x,z) = A\, e^{k_1 x-i(\omega_{1} z+\delta_{1})}\,,~~
v(x,z) = B\, e^{k_2 x-i(\omega_{2} z+\delta_{2})}\,,
\ee
provided
\be\label{3.1h}
\omega_{1}+k_{1}^2 +aA^2+bB^2 =0\,,~~\omega_{2}+k_{2}^2+fA^{2}+eB^2 = 0\,.
\ee

{\bf Solutions in terms of Lam\'e Polynomial of Order 1}

We now show that
these coupled equations admit 7 different solutions in terms of Lam\'e 
polynomials of order 1 in the $u$ and the $v$ fields, and 
3 solutions in the corresponding hyperbolic limit. Actually, there are 10
distinct solutions but since both $u$ and $v$ fields are the nonlocal NLS fields, 
the truly distinct solutions are only 7.

{\bf Solution I}

It is easily checked that 
\be\label{3.2}
u(x,t)=A \dn(\beta x,m) e^{-i (\omega_1 z+ \delta_{1})}\,,
\ee
and
\be\label{3.3}
v(x,t)=B\sqrt{m}\, \sn(\beta x,m)  e^{-i (\omega_2 z+\delta_{2})}\,,
\ee
is an exact solution to the coupled field Eqs. (\ref{3.1}) provided
\be\label{3.4}
a A^2 +b B^2 = f A^2 + eB^2 = 2\beta^2\,,
\ee
\be\label{3.5}
\omega_1 = m \beta^2 -a A^2\,,~~
\omega_2 =  (1+m) \beta^2 -f A^2\,.
\ee
On solving Eqs. (\ref{3.4}) we find that so long as $bf \ne ae$, $A,B$ are
given by
\be\label{3.6}
A^2 = \frac{2\beta^2(e-b)}{ae-bf}\,,~~B^2 = \frac{2\beta^2(a-f)}{ae-bf}\,.
\ee

Few remarks are in order at this stage. Most of these remarks apply to all
the solutions obtained below (in terms of Lam\'e polynomials of order 1).

\begin{enumerate}

\item It turns out that all the solutions in terms of Lam\'e polynomials of
order 1 are only valid if relations (\ref{3.4}) are satisfied. In case $ae=bf$, then 
along with Eqs. (\ref{3.4}) this also implies 
that $b=e, a=f$. In that case instead of the relations (\ref{3.6}), we only
have the constraint $aA^2 + b B^2 = 2\beta^2$. 

\item In the Manakov case (i.e. when $a=b=e=f$) the constraint (\ref{3.4}) becomes 
$a(A^2+B^2) = 2\beta^2$. On the other hand in the MZS case, (i.e. when $a=f=-e=-b$), 
the constraint becomes $a(A^2-B^2)=2\beta^2$. 

\end{enumerate}

{\bf Solution II}

It is easily checked that 
\be\label{3.7}
u(x,t)=A \sqrt{m}\, \cn(\beta x,m)  e^{-i (\omega_1 z+\delta_{1})}\,,
\ee
and
\be\label{3.8}
v(x,t)=B\sqrt{m}\, \sn(\beta x,m)  e^{-i (\omega_2 z+\delta_{2})}\,,
\ee
is an exact solution to the coupled field Eqs. (\ref{3.1}) provided
Eq. (\ref{3.4}) is satisfied and further
\be\label{3.9}
\omega_1 = \beta^2 -m a A^2\,,~~
\omega_2 = (1+m) \beta^2 -m f A^2 \,.
\ee

{\bf Solution III}

It is easily checked that 
\be\label{3.11}
u(x,t)=A \sqrt{m}\, \sn(\beta x,m)  e^{-i (\omega_1 z+\delta_{1})}\,,
\ee
and
\be\label{3.12}
v(x,t)=B\sqrt{m}\, \sn(\beta x,m)  e^{-i (\omega_2 z+\delta_{2})}\,,
\ee
is an exact solution to the coupled field Eqs. (\ref{3.1}) provided 
Eq. (\ref{3.4}) is satisfied and further
\be\label{3.14}
\omega_1 = \omega_2 = (1+m)\beta^2\,.
\ee

{\bf Solution IV}

It is easily checked that 
\be\label{3.16}
u(x,t)=A \sqrt{m}\, \cn(\beta x,m)  e^{-i (\omega_1 z+\delta_{1})}\,,
\ee
and
\be\label{3.17}
v(x,t)=B\sqrt{m}\, \cn(\beta x,m)  e^{-i (\omega_2 z+\delta_{2})}\,,
\ee
is an exact solution to the coupled field Eqs. (\ref{3.1}) provided
Eq. (\ref{3.4}) is satisfied and further
\be\label{3.19}
\omega_1 = \omega_2 = - (2m-1)\beta^2\,.
\ee

{\bf Solution V}

It is easily checked that 
\be\label{3.21}
u(x,t)=A \dn(\beta x,m) e^{-i (\omega_1 z+\delta_{1})}\,,
\ee
and
\be\label{3.22}
v(x,t)=B \dn(\beta x) e^{-i (\omega_2 z+\delta_{2})}\,,
\ee
is an exact solution to the coupled field Eqs. (\ref{3.1}) provided
Eq. (\ref{3.4}) is satisfied and further
\be\label{3.23}
\omega_1 = \omega_2 = - (2-m)\beta^2\,.
\ee

{\bf Solution VI}

It is easily checked that 
\be\label{3.24}
u(x,t)=A \dn(\beta x,m) e^{-i (\omega_1 z+\delta_{1})}\,,
\ee
and
\be\label{3.25}
v(x,t)=B \sqrt{m}\, \cn(\beta x,m)  e^{-i (\omega_2 z+\delta_{2})}\,,
\ee
is an exact solution to the coupled field Eqs. (\ref{3.1}) provided
Eq. (\ref{3.4}) is satisfied and further
\be\label{3.26}
\omega_1 =  - (4-3m) \beta^2-(1-m) a A^2\,,~~
\omega_2 = -(2m-1) \beta^2 -(1-m) fA^2\,.
\ee

{\bf Solution VII}

Remarkably, it turns out that a linear superposition of $\dn$ and $\cn$
is also a solution to the coupled Eqs. (\ref{3.1}). In particular, 
\be\label{3.27}
u(x,t)= \frac{1}{2} \bigg [ A\dn(\beta x,m)
+D \sqrt{m}\,  \cn(\beta x,m) \bigg ] e^{-i(\omega_{1} z+\delta_{1})}\,,
\ee
and
\be\label{3.28}
v(x,t)= \frac{1}{2} \bigg [ B \dn(\beta x,m)
+\sqrt{m}\, E \cn(\beta x,m) \bigg ]  e^{-i(\omega_{2} z+\delta_{2})}\,,
\ee
is an exact solution to the coupled field equations (\ref{3.1}) provided
Eqs. (\ref{3.4}) is satisfied and further
\be\label{3.29}
D = \pm A\,,~~E = \pm B\,,~~
\omega_1 = \omega_2 = - \frac{1}{2} (1+m)\beta^2\,.
\ee
Note that the signs of $D = \pm A$ and $E = \pm B$ are correlated.

{\bf Hyperbolic Limit}

In the limit $m=1$, all the solutions mentioned above go over to the
hyperbolic soliton solutions which we mention one by one. 

{\bf Solution VIII}

It is easily checked that in the limit $m=1$, the Solutions I and II go over
to the mixed hyperbolic (bright-dark) soliton solution
\be\label{3.30}
u(x,t)=A\, \sech(\beta x)\, e^{-i (\omega_1 z+\delta)}\,,
\ee
and
\be\label{3.31}
v(x,t)=B\, \tanh(\beta x)\, e^{-i (\omega_2 z+\delta)}\,,
\ee
provided Eq. (\ref{3.4}) is satisfied and further
\be\label{3.32}
\omega_1 =  \beta^2 -a A^2\,,~~\omega_2  = 2\beta^2 - f A^2\,.
\ee

{\bf Solution IX}

In the limit $m=1$ the Solution III goes over to the dark-dark 
soliton solution
\be\label{3.33}
u(x,t)=A\, \tanh(\beta x)\, e^{-i (\omega_1 z+\delta)}\,,
\ee
and
\be\label{3.34}
v(x,t)=B\, \tanh(\beta x)\, e^{-i (\omega_2 z+\delta)}\,,
\ee
provided Eq. (\ref{3.4}) is satisfied and further
\be\label{3.35}
\omega_1 = \omega_2 = 2 \beta^2\,.
\ee

{\bf Solution X}

Finally, in the limit $m=1$, Solutions IV to VII go over to the bright-bright
soliton solution
\be\label{3.36}
u(x,t)=A\, \sech(\beta x)\, e^{-i (\omega_1 z+\delta)}\,,
\ee
and
\be\label{3.37}
v(x,t)=B\, \sech(\beta x)\, e^{-i (\omega_2 z+\delta)}\,,
\ee
provided Eq. (\ref{3.4}) is satisfied and further
\be\label{3.38}
\omega_1 = \omega_2 = - \beta^2\,.
\ee

{\bf Solutions in Terms of Lam\'e Polynomials of Order 2}

We now show that remarkably, 
the coupled Eqs. (\ref{3.1}) admit 8 distinct periodic solutions in terms of Lam\'e 
polynomials of order 2, and 
3 solutions in the corresponding hyperbolic limit even though neither of them are 
the solutions of the uncoupled, nonlocal NLS equation. Actually, there are 17
distinct solutions but since both $u$ and $v$ fields are the nonlocal NLS fields, 
hence the truly distinct possible solutions are only 11 but it turns out that out of 
these, only 8 solutions actually exist. Remarkably, it turns out that all the
solutions in terms of Lam\'e polynomials of order two are only valid for the 
MZS case and {\it no} solution exists for the Manakov case. However, all the 
solutions also exist in the more general case of $a=f, b=e$.

{\bf Solution XI}

It is easily checked that 
\be\label{3.39}
u(x,t)=A \sqrt{m}\, \dn(\beta x,m) \cn(\beta x,m)\, e^{-i(\omega_1 z+\delta_{1})}\,,
\ee
and
\be\label{3.40}
v(x,t)=B\sqrt{m}\, \sn(\beta x,m) \dn(\beta x,m)\, e^{-i(\omega_{2} z+\delta_{2})}\,,
\ee
is an exact solution to the coupled field equations (\ref{3.1}) provided
\be\label{3.41}
a=f > 0\,,~~b=e < 0\,,~~m a A^2 = -m b B^2 = 6\beta^2\,,
\ee
\be\label{3.42}
\omega_1 = - (5-m)\beta^2\,,~~ \omega_2  = - (5-4m) \beta^2\,.
\ee
Thus while this solution will hold good in the MZS case, it 
will not hold good in the Manakov case.

{\bf Solution XII}

It is easily checked that 
\be\label{3.43}
u(x,t)=A \sqrt{m}\, \dn(\beta x,m) \cn(\beta x,m)\, e^{-i(\omega_{1} z+\delta_{1})}\,,
\ee
and
\be\label{3.44}
v(x,t)=B m\, \sn(\beta x,m) \cn(\beta x,m)\, e^{-i(\omega_{2} z+\delta_{2})}\,,
\ee
is an exact solution to the coupled field equations (\ref{3.1}) provided
Eq. (\ref{3.41}) is satisfied and further
\be\label{3.46}
\omega_1 = - (5m-1)\beta^2\,,~~ \omega_2 = - (5- 4m) \beta^2\,.
\ee

{\bf Solution XIII}

It is easily checked that 
\be\label{3.47}
u(x,t)=A m\, \cn(\beta x,m) \sn(\beta x,m)\, e^{-i(\omega_{1} z+\delta_{1})}\,,
\ee
and
\be\label{3.48}
v(x,t)=B \sqrt{m}\, \sn(\beta x,m) \dn(\beta x,m)\, e^{-i(\omega_{2} z+\delta_{2})}\,,
\ee
is an exact solution to the coupled field equations (\ref{3.1}) provided
\be\label{3.49}
a=f < 0\,,~~b=e > 0\,,~~ -(1-m) a A^2 =  (1-m) b B^2 = 6\beta^2\,,
\ee
and further
\be\label{3.50}
\omega_1 =  (4+m)\beta^2\,,~~
\omega_2  =  (1+4m) \beta^2\,.
\ee
Notice that this solution while it exists for $0 < m < 1$, it does not
hold good in the hyperbolic limit $m=1$. 

{\bf Solution XIV}

It is easily checked that 
\be\label{3.51}
u(x,t)= \big [A \dn^2(\beta x,m) +D \big ]\, e^{-i(\omega_{1} z+\delta_{1})}\,,
\ee
and
\be\label{3.52}
v(x,t)=B \sqrt{m}\, \cn(\beta x,m) \dn(\beta x,m)\, e^{-i(\omega_{2} z+\delta_{2})}\,,
\ee
is an exact solution to the coupled field equations (\ref{3.1}) provided
\be\label{3.53}
a=f\,,~~b=e\,,~~ a A^2 = - b B^2\,,~~(1-m+2z)aA^2 = 6\beta^2\,,~~
z = \frac{D}{A}\,,
\ee
\bea\label{3.54}
&&z=\frac{-(2-m) \pm \sqrt{1-m+m^2}}{3}\,, \nonumber \\
&&\omega_1-k^2 =  -[2(2-m)+3z]\beta^2\
-aA^2 z^2\,, \nonumber \\
&&\omega_2 -k^2 = - (5-m) \beta^2 -aA^2 z^2\,.
\eea

{\bf Solution XV}

It is easily checked that 
\be\label{3.55}
u(x,t)= \big [A \dn^2(\beta x,m) +D \big ]\, e^{-i(\omega_{1} z+\delta_{1})}\,,
\ee
and
\be\label{3.56}
v(x,t)=B \sqrt{m}\, \sn(\beta x,m) \dn(\beta x,m)\, e^{-i(\omega_{2} z+\delta_{2}}\,,
\ee
is an exact solution to the coupled field equations (\ref{3.1}) provided
\be\label{3.57}
a=f\,,~~b=e\,,~~ a A^2 =  -b B^2\,,~~(1+2z)aA^2 = 6\beta^2\,,~~
z = \frac{D}{A}\,,
\ee
\bea\label{3.58}
&&z=\frac{-(2-m) \pm \sqrt{1-m+m^2}}{3}\,, \nonumber \\
&&\omega_1-k^2 =  -[2(2-m)+3z]\beta^2\
-aA^2 z^2\,, \nonumber \\
&&\omega_2 -k^2 = - (5-4m) \beta^2 -aA^2 z^2\,.
\eea

{\bf Solution XVI}

It is easily checked that 
\be\label{3.59}
u(x,t)= \big [A \dn^2(\beta x,m) +D \big ]\, e^{-i(\omega_{1} z+\delta_{1})}\,,
\ee
and
\be\label{3.60}
v(x,t)=B m\, \cn(\beta x,m) \sn(\beta x,m)\, e^{-i (\omega_2 z+\delta_2)}\,, 
\ee
is an exact solution to the coupled field equations (\ref{3.1}) provided
\be\label{3.61}
a=f\,,~~b=e\,,~~ a A^2 =  -b B^2\,,~~(2-m+2z)aA^2 = 6\beta^2\,,~~
z = \frac{D}{A}\,,
\ee
\bea\label{3.62}
&&z=\frac{-(2-m) \pm \sqrt{1-m+m^2}}{3}\,, \nonumber \\
&&\omega_1-k^2 =  -[2(2-m)+3z]\beta^2\
-aA^2 [z^2-(1-m)]\,, \nonumber \\
&&\omega_2 -k^2 = - (2-m) \beta^2 -aA^2[z^2-(1-m)]\,.
\eea

{\bf Solution XVII}

It is easily checked that 
\be\label{3.63}
u(x,t)= \big [A \dn^2(\beta x,m) +D \big ]\, e^{-i (\omega_1 z+\delta_1)}\,, 
\ee
and
\be\label{3.64}
v(x,t)=  \big [B \dn^2(\beta x,m)+E \big ]\, e^{-i (\omega_2 z+\delta_2)}\,, 
\ee
is an exact solution to the coupled field equations (\ref{3.1}) provided
\be\label{3.65}
a=f\,,~~b=e\,,~~ a A^2 = - b B^2\,,~~(z_{\pm}-y_{\mp})aA^2 
= 6\beta^2\,,~~z=\frac{D}{A}\,,~~y=\frac{E}{B}\,,
\ee
\bea\label{3.66}
&&z_{\pm}=y_{\pm} =\frac{-(2-m) \pm \sqrt{1-m+m^2}}{3}\,, \nonumber \\
&&\omega_1 -k^2 = \mp 2 \sqrt{1-m+m^2} \beta^2\,,~~\omega_2 -k^2 =  
\pm 2 \sqrt{1-m+m^2} \beta^2\,.
\eea
From the relation (\ref{3.65}) it follows that this solution exists only if 
$y$ and $z$ are unequal. In addition, depending on if we choose $z_{+},y_{-}$ or 
$z_{-},y_{+}$, the corresponding $\omega_1,\omega_2$ are as given by 
Eq. (\ref{3.66}).

{\bf Solution XIII}

Remarkably, it turns out that even a supperposition of $\dn^2$ and $\cn \dn$ 
is an exact solution to the coupled Eqs. (\ref{3.1}). In particular, it is 
easily checked that 
\be\label{3.67}
u(x,t)= \bigg [\frac{A}{2} \dn^2(\beta x,m) +D + \frac{G}{2} \sqrt{m}\, 
\cn(\beta x,m) \dn(\beta x,m) \bigg ]\, e^{-i (\omega_1 z+\delta_1)}\,, 
\ee
and
\be\label{3.68}
v(x,t)= \bigg [\frac{B}{2} \dn^2(\beta x,m) +E + \frac{H}{2} \sqrt{m}\, 
\cn(\beta x,m) \dn(\beta x,m) \bigg ]\, e^{-i (\omega_2 z+\delta_2)}\,, 
\ee
is an exact solution to the coupled field equations (\ref{3.1}) provided
\bea\label{3.69}
&&G= \pm A\,,~~H = \pm B\,,~~a=f\,,~~b=e\,,~~ a A^2 = - b B^2\,,
\nonumber \\
&&(z_{\pm}-y_{\mp})aA^2 = 6\beta^2\,,~~z=\frac{D}{A}\,,~~y=\frac{E}{B}\,,
\eea
\bea\label{3.70}
&&z_{\pm}=y_{\pm} =\frac{-(5-m) \pm \sqrt{1+14m+m^2}}{12}\,, \nonumber \\
&&\omega_1 -k^2 
= \mp \frac{\sqrt{1-m+m^2}}{2} \beta^2\,,~~\omega_2 -k^2 =  
\pm \frac{\sqrt{1-m+m^2}}{2} \beta^2\,.
\eea
Thus akin to the last solution, this solution exists only if 
$y$ and $z$ are unequal. Further, depending on if we choose $z_{+},y_{-}$ or 
$z_{-},y_{+}$, the corresponding $\omega_1,\omega_2$ are as given by 
Eq. (\ref{3.70}). 

{\bf Hyperbolic Solitons of Order 2}

In the limit $m=1$, the above 8 Lam\'e polynomial solutions of order 2 reduce 
to the hyperbolic soliton solutions, which we discuss one by one.

{\bf Solution XIX}

In the limit $m=1$, the solutions XI and XII go over to the hyperbolic 
bright-dark soliton solution. In particular
\be\label{3.71}
u(x,t)=A \sech^2 (\beta x)\, e^{-i (\omega_1 z+\delta_1)}\,, 
\ee
and
\be\label{3.72}
v(x,t)=B \tanh (\beta x) \sech (\beta x)\, e^{-i (\omega_2 z+\delta_{2})}\,, 
\ee
is an exact solution to the coupled field equations (\ref{3.1}) provided
\be\label{3.73}
a=f > 0\,,~~b=e < 0\,,~~ a A^2 = -b B^2 = 6\beta^2\,,
\ee
\be\label{3.74}
\omega_1 = - 4\beta^2\,,~~ \omega_2  = - \beta^2\,.
\ee

{\bf Solution XX}

In the limit $m=1$, the solution XIV goes over to the bright-bright hyperbolic
soliton solution 
\be\label{3.75}
u(x,t)= \big [A \sech^2(\beta x) +D \big ]\, e^{-i (\omega_1 z+\delta_1)}\,,  
\ee
and
\be\label{3.76}
v(x,t)=B \sech^2 (\beta x)\, e^{-i (\omega_2 z+\delta_2)}\,, 
\ee
provided
\be\label{3.77}
a=f < 0\,,~~b=e > 0\,,~~ a A^2 = - b B^2\,,~~aA^2 = -\frac{9}{2}\beta^2\,,
\ee
\be\label{3.78}
z= \frac{D}{A} = -\frac{2}{3}\,,~~
\omega_1 =  2 \beta^2 ~~\omega_2 = -2 \beta^2\,.
\ee

{\bf Solution XXI}

In the limit $m=1$, the solutions XV and XVI go over to the bright-dark 
hyperbolic soliton solution 
\be\label{3.79}
u(x,t)= \big [A \sech^2(\beta x) +D \big ]\, e^{-i (\omega_1 z+\delta_1)}\,, 
\ee
and
\be\label{3.80}
v(x,t)=B \tanh(\beta x) \sech(\beta x)\, e^{(-i (\omega_2 z+\delta_2)}\,, 
\ee
provided
\be\label{3.81}
a=f < 0\,,~~b=e > 0\,,~~ a A^2 =  -b B^2 = - 18 \beta^2\,,
\ee
\be\label{3.82}
z= \frac{D}{A} = -\frac{2}{3}\,,~~
\omega_1 =  8 \beta^2 ~~\omega_2 = 7 \beta^2\,.
\ee

Finally, note that in the limit $m=1$, the solutions XVII and XiIII either
go over to the solution XX or solution XX with $u$ and $v$ interchanged.
Further as remarked earlier, solution XIII does not exist in the 
$m=1$ limit.

\subsection{Shifted Solutions in terms of Lam\'e Polynomials of Order 1 and 2}

We now show that this model also admits shifted solutions in terms of Lam\'e 
polynomials of order 1 and 2. 

{\bf Shifted Solutions in terms of Lam\'e Polynomials of order 1}

{\bf Solution XXII}

It is easily checked that 
\be\label{3.83}
u(x,t)=A \sqrt{m} \frac{\cn(\beta x,m)}{\dn(\beta x,m)} 
e^{-i (\omega_1 z+ \delta_{1})}\,,
\ee
and
\be\label{3.84}
v(x,t)=B\sqrt{m(1-m)}\, \frac{\sn(\beta x,m)}{\dn(\beta x,m)}  
e^{-i (\omega_2 z+\delta_{2})}\,,
\ee
is an exact solution to the coupled field Eqs. (\ref{3.1}) provided
\be\label{3.85}
a A^2 +b B^2 = f A^2 + eB^2 = -2\beta^2\,,
\ee
\be\label{3.86}
0< m < 1\,,~~\omega_1 = (1-m) \beta^2 -m a A^2\,,~~
\omega_2 =  -(2m-1) \beta^2 -m f A^2\,.
\ee
On solving Eqs. (\ref{3.4}) we find that so long as $bf \ne ae$, $A,B$ are
given by
\be\label{3.87}
A^2 = \frac{2\beta^2(b-e)}{ae-bf}\,,~~B^2 = \frac{2\beta^2(f-a)}{ae-bf}\,.
\ee

Few remarks are in order at this stage. These remarks also apply to 
the next two solutions and hence we 
will not repeat them while discussing these two solutions.

\begin{enumerate}

\item It turns out that the solutions XXII, XXIII and XXIV in terms of Lam\'e
 polynomial of order 1 are only valid if relations (\ref{3.85}) are satisfied. In case $ae=bf$, then along with Eqs. (\ref{3.85}) this also implies 
that $b=e, a=f$. In that case instead of the relations (\ref{3.87}), we only
have the constraint $aA^2 + b B^2 = -2\beta^2$. 

\item In the Manakov case (i.e. when $a=b=e=f$) the constraint (\ref{3.4}) becomes 
$a(A^2+B^2) = -2\beta^2$. On the other hand in the MZS case, (i.e. when $a=f=-e=-b$), 
the constraint becomes $a(A^2-B^2)= -2\beta^2$. 

\end{enumerate}
 
{\bf Solution XXIII}

It is easily checked that 
\be\label{3.88}
u(x,t)=A \sqrt{m} \frac{\cn(\beta x,m)}{\dn(\beta x,m)} 
e^{-i (\omega_1 z+ \delta_{1})}\,,
\ee
and
\be\label{3.89}
v(x,t)=B\sqrt{m}\, \frac{\cn(\beta x,m)}{\dn(\beta x,m)}  
e^{-i (\omega_2 z+\delta_{2})}\,,
\ee
is an exact solution to the coupled field Eqs. (\ref{3.1}) provided
relations (\ref{3.85}) are satisfied and further
\be\label{3.91}
\omega_1 = \omega_{2} = (1+m) \beta^2\,.
\ee

{\bf Solution XXIV}

It is easily checked that 
\be\label{3.92}
u(x,t)=A \sqrt{m(1-m)} \frac{\sn(\beta x,m)}{\dn(\beta x,m)} 
e^{-i (\omega_1 z+ \delta_{1})}\,,
\ee
and
\be\label{3.93}
v(x,t)=B\sqrt{m(1-m)}\, \frac{\sn(\beta x,m)}{\dn(\beta x,m)}  
e^{-i (\omega_2 z+\delta_{2})}\,,
\ee
is an exact solution to the coupled field Eqs. (\ref{3.1}) provided
relations (\ref{3.85}) are satisfied and further
\be\label{3.94}
\omega_1 = \omega_{2} = -(2m-1) \beta^2\,.
\ee

{\bf Solution XXV}

It is easily checked that 
\be\label{3.95}
u(x,t)=A \sqrt{m} \frac{\cn(\beta x,m)}{\dn(\beta x,m)} 
e^{-i (\omega_1 z+ \delta_{1})}\,,
\ee
and
\be\label{3.96}
v(x,t)=B\sqrt{(1-m)}\, \frac{1}{\dn(\beta x,m)}  
e^{-i (\omega_2 z+\delta_{2})}\,,
\ee
is an exact solution to the coupled field Eqs. (\ref{3.1}) provided
\be\label{3.97}
b B^2- a A^2 = e B^2 -f A^2  = 2\beta^2\,,
\ee
\be\label{3.98}
0< m < 1\,,~~\omega_1 = -(1-m) \beta^2 - a A^2\,,~~
\omega_2 =  -(2-m) \beta^2 - f A^2\,.
\ee
On solving Eqs. (\ref{3.97}) we find that so long as $bf \ne ae$, $A,B$ are
given by
\be\label{3.99}
A^2 = \frac{2\beta^2(b-e)}{ae-bf}\,,~~B^2 = \frac{2\beta^2(a-f)}{ae-bf}\,.
\ee
However, in case $ae=bf$ then $a=f,b=e$ and then instead of the relations 
(\ref{3.99}), we only have the constraint $2\beta^2 = b B^2 - a A^2$. In the
Manakov limit (i.e. $a=f=b=e$), we have the constraint $a(A^2-B^2) 
= -2 \beta^2$ while in the MZS case (i.e. $a=f=-b=-e$) we have the constraint
$a(A^2+B^2) = -2\beta^2$. 

{\bf Solution XXVI}

It is easily checked that 
\be\label{3.100}
u(x,t)=A \sqrt{m(1-m)} \frac{\sn(\beta x,m)}{\dn(\beta x,m)} 
e^{-i (\omega_1 z+ \delta_{1})}\,,
\ee
and
\be\label{3.101}
v(x,t)=B\sqrt{(1-m)}\, \frac{1}{\dn(\beta x,m)}  
e^{-i (\omega_2 z+\delta_{2})}\,,
\ee
is an exact solution to the coupled field Eqs. (\ref{3.1}) provided
relations (\ref{3.97}) are satisfied and further 
\be\label{3.102}
0< m < 1\,,~~\omega_1 = -\beta^2 -(1-m) a A^2\,,~~
\omega_2 =  -(2-m) \beta^2 - (1-m) f A^2\,.
\ee
Thus all the remarks made after the previous solution (i.e. solution XXV) are
also valid in this case.  

{\bf Solution XXVII}

It is easily checked that 
\be\label{3.103}
u(x,t)=A \sqrt{(1-m)} \frac{1}{\dn(\beta x,m)} 
e^{-i (\omega_1 z+ \delta_{1})}\,,
\ee
and
\be\label{3.104}
v(x,t)=B\sqrt{(1-m)}\, \frac{1}{\dn(\beta x,m)}  
e^{-i (\omega_2 z+\delta_{2})}\,,
\ee
is an exact solution to the coupled field Eqs. (\ref{3.1}) provided
\be\label{3.105}
a A^2 + b B^2 = f A^2 + e B^2  = 2\beta^2\,,
\ee
\be\label{3.106}
0< m < 1\,,~~\omega_1 = \omega_{2} = -(2-m) \beta^2\,.
\ee
On solving Eqs. (\ref{3.105}) we find that so long as $bf \ne ae$, $A,B$ are
given by
\be\label{3.107}
A^2 = \frac{2\beta^2(e-b)}{ae-bf}\,,~~B^2 = \frac{2\beta^2(a-f)}{ae-bf}\,.
\ee
However, in case $ae=bf$ then $a=f,b=e$ and then instead of the relations 
(\ref{3.107}), we only have the constraint $2\beta^2 = a A^2 + b B^2$. In the
Manakov limit (i.e. $a=f=b=e$), we have the constraint $a(A^2+B^2) 
= 2 \beta^2$ while in the MZS case (i.e. $a=f=-b=-e$) we have the constraint
$a(A^2-B^2) = 2\beta^2$. 

Summarizing, we have presented 6 shifted solutions in terms of Lam\'e polynomials 
of order 1, all of which are only valid in case $0 < m < 1$ but not at $m=1$.

{\bf Shifted Solutions in Terms of Lam\'e Polynomials of order 2}

We now present 7 solutions of the coupled Eqs. (\ref{3.1}) in terms of shifted
Lam\'e polynomials of order 2, even though none of them is a solution of the
corresponding uncoupled problem.

{\bf Solution XXVIII}

It is easily checked that 
\be\label{3.108}
u(x,t)=A \sqrt{m} \frac{\cn(\beta x,m)}{\dn^{2}(\beta x,m)} 
e^{-i (\omega_1 z+ \delta_{1})}\,,
\ee
and
\be\label{3.109}
v(x,t)=B\sqrt{m(1-m)}\, \frac{\sn(\beta x,m)}{\dn^{2}(\beta x,m)}  
e^{-i (\omega_2 z+\delta_{2})}\,,
\ee
is an exact solution to the coupled field Eqs. (\ref{3.1}) provided
\be\label{3.110}
a=f > 0\,,~~b=e < 0\,,~~a A^2 = - b B^2\,,~~a m A^2 = 6(1-m) \beta^2\,,
\ee
\be\label{3.111}
0< m < 1\,,~~\omega_1 = -(5-4m) \beta^2\,,~~
\omega_{2} = -(5-m) \beta^2\,.
\ee
Thus this solution can only be valid in the MZS case but not in the Manakov case.

{\bf Solution XXIX}

It is easily checked that 
\be\label{3.112}
u(x,t)=A \sqrt{m} \frac{\cn(\beta x,m)}{\dn^{2}(\beta x,m)} 
e^{-i (\omega_1 z+ \delta_{1})}\,,
\ee
and
\be\label{3.113}
v(x,t)=B m\, \frac{\cn(\beta x,m) \sn(\beta x,m)}{\dn^{2}(\beta x,m)}  
e^{-i (\omega_2 z+\delta_{2})}\,,
\ee
is an exact solution to the coupled field Eqs. (\ref{3.1}) provided
\be\label{3.114}
a=f < 0\,,~~b=e < 0\,,~~a A^2 = b B^2\,,~~a A^2 = -6 \beta^2\,,
\ee
\be\label{3.115}
0< m < 1\,,~~\omega_1 = (4m+1) \beta^2\,,~~
\omega_{2} = (4+m) \beta^2\,.
\ee
Thus this solution can only be valid in the Manakov case but not in the MZS case.

{\bf Solution XXX}

It is easily checked that 
\be\label{3.116}
u(x,t)=A \sqrt{m(1-m)} \frac{\sn(\beta x,m)}{\dn^{2}(\beta x,m)} 
e^{-i (\omega_1 z+ \delta_{1})}\,,
\ee
and
\be\label{3.117}
v(x,t)=B m\, \frac{\cn(\beta x,m) \sn(\beta x,m)}{\dn^{2}(\beta x,m)}  
e^{-i (\omega_2 z+\delta_{2})}\,,
\ee
is an exact solution to the coupled field Eqs. (\ref{3.1}) provided
\be\label{3.118}
a=f < 0\,,~~b=e < 0\,,~~a A^2 = b B^2\,,~~a A^2 = -6(1-m) \beta^2\,,
\ee
\be\label{3.119}
0< m < 1\,,~~\omega_1 = -(5m-1) \beta^2\,,~~
\omega_{2} = -(5m-4) \beta^2\,.
\ee
Thus this solution can only be valid in the Manakov case but not in the MZS case.

{\bf Solution XXXI}

It is easily checked that 
\be\label{3.120}
u(x,t)=A \sqrt{m} \frac{\cn(\beta x,m)}{\dn^{2}(\beta x,m)} 
e^{-i (\omega_1 z+ \delta_{1})}\,,
\ee
and
\be\label{3.121}
v(x,t)= \left[\frac{B}{\dn^{2}(\beta x,m)}+D\right]  
e^{-i (\omega_2 z+\delta_{2})}\,,
\ee
is an exact solution to the coupled field Eqs. (\ref{3.1}) provided
\be\label{3.122}
a=f\,,~~b=e\,,~~a A^2 = b B^2\,,~~a A^2[1+2(1-m)y] = 6(1-m) \beta^2\,,
\ee
\be\label{3.123}
0< m < 1\,,~~\omega_1 = -(5m-4) \beta^2 -(1-m)a A^2 y^2\,,~~
\omega_{2} = -2[2(2-m)+3(1-m)x] \beta^2-(1-m) a A^2 y^2\,,
\ee
where
\be\label{3.124}
y = \frac{D}{B} = -\frac{(2-m)}{3(1-m)} \pm \frac{\sqrt{1-m+m^2}}{3(1-m)}\,.
\ee
Thus this solution can only be valid in the Manakov case but not in the 
MZS case.

{\bf Solution XXXII}

It is easily checked that 
\be\label{3.125}
u(x,t)=A \sqrt{m(1-m)} \frac{\sn(\beta x,m)}{\dn^{2}(\beta x,m)} 
e^{-i (\omega_1 z+ \delta_{1})}\,,
\ee
and
\be\label{3.126}
v(x,t)= \left[\frac{B}{\dn^{2}(\beta x,m)}+D\right]  
e^{-i (\omega_2 z+\delta_{2})}\,,
\ee
is an exact solution to the coupled field Eqs. (\ref{3.1}) provided
Eq. (\ref{3.122}) is satisfied with $y$ given by Eq. (\ref{3.124}) and further
\be\label{3.127}
0< m < 1\,,~~\omega_1 = -(5-m) \beta^2 -(1-m)a A^2 y^2\,,~~
\omega_{2} = -2[2(2-m)+3(1-m)y] \beta^2-(1-m) a A^2 y^2\,.
\ee
Thus this solution can only be valid in the Manakov case but not in the MZS case.

{\bf Solution XXXIII}

It is easily checked that 
\be\label{3.128}
u(x,t)=A m \frac{\cn(\beta x,m) \sn(\beta x,m)}{\dn^{2}(\beta x,m)} 
e^{-i (\omega_1 z+ \delta_{1})}\,,
\ee
and
\be\label{3.129}
v(x,t)= \left[\frac{B}{\dn^{2}(\beta x,m)}+D\right]  
e^{-i (\omega_2 z+\delta_{2})}\,,
\ee
is an exact solution to the coupled field Eqs. (\ref{3.1}) provided
\be\label{3.130}
a=f\,,~~b=e\,,~~a A^2 = - b B^2\,,~~[2-m+ 2(1-m)y] a A^2 = -6(1-m) \beta^2\,,
\ee
\be\label{3.131}
0< m < 1\,,~~\omega_1 = -(2-m) \beta^2 -[1-(1-m) y^2]a A^2\,,~~
\omega_{2} = -2[2(2-m)+3(1-m)y] \beta^2-[1-(1-m)y^2] a A^2\,,
\ee
with $y$ given by Eq. (\ref{3.124}). 

Thus this solution can only be valid in the MZS case but not in the Manakov case.

{\bf Solution XXXIV}

Finally, it is easily checked that 
\be\label{3.132}
u(x,t)=\left[\frac{A}{\dn^{2}(\beta x,m)}+E\right] 
e^{-i (\omega_1 z+ \delta_{1})}\,,
\ee
and
\be\label{3.133}
v(x,t)= \left[\frac{B}{\dn^{2}(\beta x,m)}+D\right]  
e^{-i (\omega_2 z+\delta_{2})}\,,
\ee
is an exact solution to the coupled field Eqs. (\ref{3.1}) provided
\be\label{3.134}
a=f\,,~~b=e\,,~~a A^2 = - b B^2\,,~~y \ne z\,,~~(y_{\pm} -z_{\mp}) a A^2 = 3 \beta^2\,,
\ee
\be\label{3.135}
0< m < 1\,,~~\omega_1 = -[4(2-m)+3(1-m)(3y+z)] \beta^2\,,~~
\omega_{2} = -[4(2-m)+3(1-m)(3z+y)] \beta^2\,,
\ee
with $y \ne z$ and both given by Eq. (\ref{3.124}). 

Thus this solution can only be valid in the MZS case but not in the Manakov case.

Summarizing, there are 7 periodic solutions in terms of shifted Lam\'e 
polynomials of order two in case $0 < m < 1$ 
out of which 4 solutions can only exist in the Manakov case (but not in the MZS
case) while 3 solutions can only exist in the MZS case (but not in the Manakov case).

\section{Nonlocal, Discrete, Saturable  NLSE}

Motivated by the nonlocal NLSE discussed in I and II, we now introduce a nonlocal
saturable DNLSE given by
\be\label{16}
i\frac{du_{n}}{dz}(z) +[u_{n+1}(z)+u_{n-1}(z)-2u_{n}(z)]
+ \frac{g u_{n}(z) u^{*}_{-n}(z)}{1+u_{n}(z)u^{*}_{-n}(z)} u_{n}(z)  =0\,,
\ee
which describes a non-Hermitian but PT-invariant system. In this case the power
as given by 
\be\label{16a}
P = \sum_{i= -\infty}^{\infty} |u_{i}|^2\,,
\ee  
is not conserved while the quasi-power 
\be\label{17}
Q = \sum_{i= -\infty}^{\infty} u_{n}(z)u^{*}_{-n}(z)\,,
\ee
is conserved. The nonlocal, discrete, saturable NLSE given by Eq. (\ref{16}) is 
a non-Hermitian but PT-invariant nonlinear system in the sense that
$V_{n}(z) = V^{*}_{-n}(z)$ where $V_{n}(z) =u_{n}(z)u^{*}_{-n}(z)$. 

Before we obtain the periodic solutions to the discrete Eq. (\ref{16}), we remark
that Eq. (\ref{16}) has a novel solution
\be\label{17a}
u_{n}(z) = A\, e^{kn -i\omega(z+\delta)}
\ee
provided
\be\label{17b}
A^2 = \frac{4\sin^{2}(\frac{k}{2})-\omega}{g+\omega-4\sin^{2}(\frac{k}{2})}\,.
\ee

Let us now obtain discrete periodic soliton solutions to this equation. In order
to obtain results in this and the next section, we have used a number of (not so
well known) local identities for Jacobi elliptic functions \cite{kls}

{\bf Solution I}

It is easily checked that one of the exact periodic solution to this equation is
\be\label{18}
u_{n} = A \dn(\beta n,m)\, e^{-i(\omega z+\delta)}\,,
\ee
provided
\be\label{19}
A^2 \cs^{2}(\beta,m)= 1\,,~~\omega = (2-g)= 
2 \left [1-\frac{\dn(\beta,m)}{\cn^{2}(\beta,m)} \right ]\,.
\ee

{\bf Solution II}

Another exact periodic solution to the discrete, nonlocal, saturable NLS 
Eq. (\ref{16}) is
\be\label{20}
u_{n} = A \sqrt{m}\, \cn(\beta n,m)\, e^{-i(\omega z+\delta)}\,,
\ee
provided
\be\label{21}
A^2 \ds^{2}(\beta,m)= 1\,,~~\omega = (2-g)= 
2 \left [1-\frac{\cn(\beta,m)}{\dn^{2}(\beta,m)}\right ]\,.
\ee

{\bf Solution III}

Remarkably, even a linear superposition of the two, i.e.
\be\label{22}
u_{n} = \left [ \frac{A}{2} \dn(\beta n,m)
+ \frac{B}{2} \sqrt{m}\, \cn(\beta n,m) \right ]\, e^{-i(\omega z+\delta)}\,,
\ee
is also an exact periodic solution to this equation provided
\be\label{23}
B = \pm A\,,~~A^2 [\cs(\beta,m)+\ds(\beta,m)]^2 = 4\,,~~\omega = (2-g)= 
2\left [1-\frac{2}{\cn(\beta,m)+\dn(\beta,m)}\right ]\,.
\ee
Here 
\be\label{24}
\cs(\beta,m) =\frac{\cn(\beta,m)}{\sn(\beta,m)}\,,~~
\ds(\beta,m) =\frac{\dn(\beta,m)}{\sn(\beta,m)}\,,~~
\ns(\beta,m) =\frac{1}{\sn(\beta,m)}\,.
\ee

{\bf Solution IV}

In the limit $m=1$, all three periodic solutions, i.e. $\cn,\dn$ as well
as the superposed solution with $B=+A$ go over to the same hyperbolic bright 
soliton solution 
\be\label{25}
u_{n} = A \sech(\beta n)\, e^{-i(\omega z+\delta)}\,,
\ee
provided
\be\label{26}
A^2 = \sech^{2}(\beta,m)\,,~~\omega=2-g = -2[\cosh(\beta,m)-1]\,,
\ee
while the solution with $B=-A$ goes over to the vacuum solution. 

{\bf Solution V}

Remarkably, unlike the local saturable discrete NLSE \cite{krss}, the  nonlocal, 
saturable, discrete  NLSE (\ref{16}) (which admits both $\dn$ and $\cn$ solutions) 
also admits the periodic $\sn$ solution
\be\label{27}
u_{n} = A \sqrt{m}\, \sn(\beta n,m)\, e^{-i(\omega z+\delta)}\,,
\ee
provided
\be\label{28}
A^2 \ns^{2}(\beta,m)= 1\,,~~\omega = (2-g)= 
2[1- \cn(\beta,m) \dn(\beta,m)]\,.
\ee

{\bf Solution VI}

In the limit $m=1$, this periodic solution goes over to the hyperbolic dark
soliton solution
\be\label{29}
u_{n} = A \tanh(\beta n) e^{-i(\omega z+\delta)}\,,
\ee
provided
\be\label{30}
A^2 = \tanh^{2}(\beta)\,,~~\omega = (2-g) = 2 \tanh^{2}(\beta)\,.
\ee

\subsection{Shifted Periodic Solutions}

We now show that this model also admits shifted periodic solutions which we now
discuss one by one. 

{\bf Solution VII}

It is easily checked that
\be\label{31}
u_{n} = \frac{A}{\dn(n \beta,m)}\,e^{-i(\omega z+\delta)}\,,
\ee
is an exact solution to Eq. (\ref{16}) provided
\be\label{32}
A^2 = \frac{[\cs(\beta,m) \cn^{2}(\beta,m)-2\cs(2\beta,m)\dn(\beta,m)]}
{\cn^{2}(\beta,m)}\,,~~\omega =2-g = -\frac{\dn(\beta,m)}{\cn^{2}(\beta,m)}\,.
\ee

{\bf Solution VIII}

It is easily checked that
\be\label{33}
u_{n} = A \sqrt{m} \frac{\cn(n \beta,m)}{\dn(n \beta,m)}\,e^{-i(\omega z+\delta)}\,,
\ee
is an exact solution to Eq. (\ref{16}) provided
\bea\label{34}
&&g=(\omega-2) \cs^{2}(\beta,m) =-2(1+ A^2)\, \ns(\beta,m) [\cs(2\beta,m)+\ds(2\beta,m)]\,,
\nonumber \\
&&(\omega-2) \cs(\beta,m)\cs(2\beta,m) \ds(\beta,m) = -\cs^{2}(\beta,m)
[\ds(2\beta,m)+\cs(2\beta,m)] \nonumber \\
&&- A^2\, \ds^{2}(\beta,m) [\cs(2\beta,m)+\ds(2\beta,m)]\,.
\eea
On solving these two equations, one can obtain $\omega$ and $gA^2$.

{\bf Solution IX}

It is easily checked that
\be\label{35}
u_{n} = A \sqrt{m} \frac{\sn(n \beta,m)}{\dn(n \beta,m)}\,e^{-i(\omega z+\delta)}\,,
\ee
is an exact solution to Eq. (\ref{16}) provided
\bea\label{36}
&&g = (\omega-2)  \cs^{2}(\beta,m) 
=-2(1+ A^2)\, \ds(\beta,m) [\cs(2\beta,m)+\ns(2\beta,m)]\,,
\nonumber \\
&&(\omega-2) \cs(\beta,m)\cs(2\beta,m) \ns(\beta,m) = -\cs^{2}(\beta,m)
[\ns(2\beta,m)+\cs(2\beta,m)] \nonumber \\
&& - A^2\, \ns^{2}(\beta,m) [\cs(2\beta,m)+\ds(2\beta,m)]\,.
\eea
On solving these two equations, one can obtain $\omega$ and $gA^2$.

\section{Coupled Nonlocal Saturable DNLSE}

We now consider a coupled nonlocal saturable DNLSE which is similar but
more general than the one considered by us previously \cite{ks6} 
\be\label{5.1}
i\frac{du_{n}}{dz}(z) +[u_{n+1}(z)+u_{n-1}(z)-2u_{n}(z)]
+g_{1} \frac{a u_{n}(z) u^{*}_{-n}(z)+b v_{n}(z) v^{*}_{-n}(z)}
{1+a u_{n}(z)u^{*}_{-n}(z)+b v_{n}(z) v^{*}_{-n}(z)} u_{n}(z)  =0\,,
\ee
\be\label{5.2}
i\frac{dv_{n}}{dz}(z) +[v_{n+1}(z)+v_{n-1}(z)-2v_{n}(z)]
+g_{2} \frac{d u_{n}(z) u^{*}_{-n}(z)+e v_{n}(z) v^{*}_{-n}(z)}
{1+d u_{n}(z)u^{*}_{-n}(z)+e v_{n}(z) v^{*}_{-n}(z)} v_{n}(z)  =0\,.
\ee
It is easy to check that these coupled nonlocal equations will have several
solutions. As an illustration, we now present a few such  solutions.

{\bf Solution I}

It is easy to check that 
\be\label{5.3}
u_{n} = A \dn(n \beta,m)\,e^{-i(\omega_{1} z+\delta)}\,,
\ee
\be\label{5.4}
v_{n} = B \sqrt{m} \sn(n \beta,m)\,e^{-i(\omega_{2} z+\delta)}\,,
\ee
is an exact solution to the coupled Eqs. (\ref{5.1}) and (\ref{5.2})
provided
\be\label{5.5}
aA^2 \cs^{2}(\beta,m) +b B^2  \ns^{2}(\beta,m) =
dA^2 \cs^{2}(\beta,m) +e B^2  \ns^{2}(\beta,m)= 1\,, 
\ee
\be\label{5.6}
g_1 = 2-\omega_{1} =  2(1- b B^2) \frac{\dn(\beta,m)}{\cn^{2}(\beta,m)}\,,
~~g_2 = 2-\omega_{2} = 2 (1-d A^2)  \cn(\beta,m) \dn(\beta,m)\,.
\ee
On solving relations (\ref{5.5}) we find that so long as $ae \ne bd$
\be\label{5.7}
A^2 \cs^{2}(\beta,m) = \frac{e-b}{ae-bd}\,,~~B^2 \ns^{2}(\beta) 
= \frac{a-d}{ae-bd}\,.
\ee
However, in case $ae =bd$, this implies that $a=d,b=e$ and instead of relations
(\ref{5.7}), we only have the constraint 
$aA^2 \cs^{2}(\beta,m)+bB^2 \ns^{2}(\beta,m) =1$. 

{\bf Solution II}

It is easy to check that 
\be\label{5.8}
u_{n} = A \sqrt{m} \cn(n \beta,m)\,e^{-i(\omega_{1} z+\delta)}\,,
\ee
\be\label{5.9}
v_{n} = B \sqrt{m} \sn(n \beta,m)\,e^{-i(\omega_{2} z+\delta)}\,,
\ee
is an exact solution to the coupled Eqs. (\ref{5.1}) and (\ref{5.2})
provided
\be\label{5.10}
aA^2 \ds^{2}(\beta,m) +b B^2  \ns^{2}(\beta,m) = 
d A^2 \ds^2(\beta,m) +eB^2 \ns^{2}(\beta,m) =1\,,
\ee
\be\label{5.11}
g_1 = 2-\omega_{1} = 2 (1-b B^2 m) \frac{\cn(\beta,m)}{\dn^{2}(\beta,m)}\,,
~~g_2 = 2-\omega_{2} = 2 (1+d A^2 m) \cn(\beta,m) \dn(\beta,m)\,.
\ee
On solving relations (\ref{5.10}) we find that so long as $ae \ne bd$
\be\label{5.12}
A^2 \ds^{2}(\beta,m) = \frac{e-b}{ae-bd}\,,~~B^2 \ns^{2}(\beta) 
= \frac{a-d}{ae-bd}\,.
\ee
However, in case $ae =bd$, this implies that $a=d,b=e$ and instead of relations
(\ref{5.12}), we only have the constraint 
$aA^2 \cs^{2}(\beta,m)+bB^2 \ns^{2}(\beta,m) =1$. 

{\bf Solution III}

It is easy to check that 
\be\label{5.13}
u_{n} = A \dn(n \beta,m)\,e^{-i(\omega_{1} z+\delta)}\,,
\ee
\be\label{5.14}
v_{n} = B \dn(n \beta,m)\,e^{-i(\omega_{2} z+\delta)}\,,
\ee
is an exact solution to the coupled Eqs. (\ref{5.1}) and (\ref{5.2})
provided
\be\label{5.15}
(aA^2+bB^2) \cs^{2}(\beta,m) =
(dA^2+eB^2) \cs^{2}(\beta,m) = 1\,, 
\ee
\be\label{5.16}
\omega_{1} = \omega_{2} =  2\left[1- \frac{\dn(\beta,m)}{\cn^{2}(\beta,m)}\right]\,,
~~g_1 = g_2 = 2- \omega_{1}\,.
\ee
On solving relations (\ref{5.15}) we find that so long as $ae \ne bd$
\be\label{5.17}
A^2 \cs^{2}(\beta,m) = \frac{e-b}{ae-bd}\,,~~B^2 \cs^{2}(\beta) 
= \frac{a-d}{ae-bd}\,.
\ee
However, in case $ae =bd$, this implies that $a=d,b=e$ and instead of relations
(\ref{5.17}), we only have the constraint 
$(aA^2+bB^2) \cs^{2}(\beta,m) =1$. 

{\bf Solution IV}

It is easy to check that 
\be\label{5.18}
u_{n} = A \sqrt{m} \cn(n \beta,m)\,e^{-i(\omega_{1} z+\delta)}\,,
\ee
\be\label{5.19}
v_{n} = B \sqrt{m} \cn(n \beta,m)\,e^{-i(\omega_{2} z+\delta)}\,,
\ee
is an exact solution to the coupled Eqs. (\ref{5.1}) and (\ref{5.2})
provided
\be\label{5.20}
(aA^2+bB^2) \ds^{2}(\beta,m) =
(dA^2+eB^2) \ds^{2}(\beta,m) = 1\,, 
\ee
\be\label{5.21}
\omega_{1} = \omega_{2} =  2\left[1- \frac{\cn(\beta,m)}{\dn^{2}(\beta,m)}\right]\,,
~~g_1 = g_2 = 2-\omega_{1}\,.
\ee
On solving relations (\ref{5.20}) we find that so long as $ae \ne bd$
\be\label{5.22}
A^2 \ds^{2}(\beta,m) = \frac{e-b}{ae-bd}\,,~~B^2 \ds^{2}(\beta) 
= \frac{a-d}{ae-bd}\,.
\ee
However, in case $ae =bd$, this implies that $a=d,b=e$ and instead of relations
(\ref{5.22}), we only have the constraint 
$(aA^2+bB^2) \ds^{2}(\beta,m) =1$. 

{\bf Solution V}

It is easy to check that 
\be\label{5.23}
u_{n} = A \dn(n \beta,m)\,e^{-i(\omega_{1} z+\delta)}\,,
\ee
\be\label{5.24}
v_{n} = B \sqrt{m} \cn(n \beta,m)\,e^{-i(\omega_{2} z+\delta)}\,,
\ee
is an exact solution to the coupled Eqs. (\ref{5.1}) and (\ref{5.2})
provided
\be\label{5.25}
aA^2 \cs^{2}(\beta,m) +bB^2 \ds^{2}(\beta,m) =
dA^2 \cs^{2}(\beta,m) +eB^2 \ds^{2}(\beta,m) = 1\,, 
\ee
\be\label{5.26}
g_1 = 2-\omega_{1} = 2[1-(1-m) bB^2]  \frac{\dn(\beta,m)}{\cn^{2}(\beta,m)}\,,
~~g_2 = 2-\omega_{2} = 2[1+ (1-m) d A^2]  
\frac{\cn(\beta,m)}{\dn^{2}(\beta,m)}\,.
\ee
On solving relations (\ref{5.25}) we find that so long as $ae \ne bd$
\be\label{5.27}
A^2 \ds^{2}(\beta,m) = \frac{e-b}{ae-bd}\,,~~B^2 \ds^{2}(\beta) 
= \frac{a-d}{ae-bd}\,.
\ee
However, in case $ae =bd$, this implies that $a=d,b=e$ and instead of relations
(\ref{5.27}), we only have the constraint 
$(aA^2+bB^2) \ds^{2}(\beta,m) =1$. 

{\bf Solution VI}

It is easy to check that 
\be\label{5.28}
u_{n} = A \sqrt{m} \sn(n \beta,m)\,e^{-i(\omega_{1} z+\delta)}\,,
\ee
\be\label{5.29}
v_{n} = B \sqrt{m} \sn(n \beta,m)\,e^{-i(\omega_{2} z+\delta)}\,,
\ee
is an exact solution to the coupled Eqs. (\ref{5.1}) and (\ref{5.2})
provided
\be\label{5.30}
(aA^2+bB^2) \ns^{2}(\beta,m) =
(dA^2+eB^2) \ns^{2}(\beta,m) = 1\,, 
\ee
\be\label{5.31}
\omega_{1} = \omega_{2} =  2[1- \cn(\beta,m) \dn (\beta,m)]\,,
~~g_1 = g_2 = 2-\omega_{1}\,.
\ee
On solving relations (\ref{5.30}) we find that so long as $ae \ne bd$
\be\label{5.32}
A^2 \ns^{2}(\beta,m) = \frac{e-b}{ae-bd}\,,~~B^2 \ns^{2}(\beta) 
= \frac{a-d}{ae-bd}\,.
\ee
However, in case $ae =bd$, this implies that $a=d,b=e$ and instead of relations
(\ref{5.32}), we only have the constraint 
$(aA^2+bB^2) \ns^{2}(\beta,m) =1$. 

In the limit $m=1$ these coupled solutions go over to the corresponding hyperbolic 
pulse or kink solutions which can be easily worked out from here.

\section{Nonlocal Ablowitz-Ladik Equation}

Yet another nonlocal nonlinear discrete equation that we now consider is 
the nonlocal AL equation
\be\label{37}
i\frac{du_{n}}{dz}(z) + u_{n+1}(z)+u_{n-1}(z)
+ g u_{n}(z) u^{*}_{-n}(z) [u_{n+1}(z)+u_{n-1}(z)] = 0\,.
\ee
Note that this is also a non-Hermitian, but PT-invariant 
system. In this case, the power as given by 
\be\label{38}
P = \sum_{i=-\infty}^{\infty} \ln[1+gu_{n}(z)u^{*}_{n}(z)]
\ee
is not conserved while the quasi-power as given by 
\be\label{39}
Q = \sum_{i=-\infty}^{\infty} \ln[1+gu_{n}(z)u^{*}_{-n}(z)]
\ee
is conserved. Note that the 
corresponding local AL equation is a well known discrete integrable equation 
\cite{al}. 

Before obtaining periodic solutions, we note that there is a novel exact solution to
the nonlocal AL Eq. (\ref{37}) given by
\be\label{40}
u_{n}(z) = A\, e^{kn-i(\omega z +\delta)}\,,
\ee
provided 
\be\label{41}
gA^2 = -\frac{\omega+2\cos(k)}{2\cos(k)}\,.
\ee

Let us now obtain discrete periodic soliton solutions to this equation. 

{\bf Solution I}

It is easily checked that one of the exact periodic solution to this equation is
\be\label{42}
u_{n} = A \dn(\beta n,m)\, e^{-i(\omega z+\delta)}\,,
\ee
provided
\be\label{43}
g A^2 \cs^{2}(\beta,m)= 1\,,~~\omega =  
-2\frac{\dn(\beta,m)}{\cn^{2}(\beta,m)}\,.
\ee

{\bf Solution II}

Another exact periodic solution to the nonlocal AL Eq. (\ref{37}) is
\be\label{44}
u_{n} = A \sqrt{m}\, \cn(\beta n,m)\, e^{-i(\omega z+\delta)}
\ee
provided
\be\label{45}
g A^2 \ds^{2}(\beta,m)= 1\,,~~\omega = 
-2\frac{\cn(\beta,m)}{\dn^{2}(\beta,m)}\,.
\ee

{\bf Solution III}

Remarkably, even a linear superposition of the two, i.e.
\be\label{46}
u_{n} = \left [ \frac{A}{2} \dn(\beta n,m)
+ \frac{B}{2} \sqrt{m}\, \cn(\beta n,m) \right ]\, e^{-i(\omega z+\delta)}\,,
\ee
is also an exact periodic solution to this equation provided
\be\label{47}
B = \pm A\,,~~g A^2 [\cs(\beta,m)+\ds(\beta,m)]^2 = 4\,,~~\omega =  
-\frac{4}{\cn(\beta,m)+\dn(\beta,m)}\,.
\ee

{\bf Solution IV}

In the limit $m=1$, all three periodic solutions, i.e. $\cn,\dn$ as well
as the superposed solution with $B=+A$ go over to the same hyperbolic pulse 
solution 
\be\label{48}
u_{n} = A \sech(\beta n)\, e^{-i(\omega z+\delta)}\,,
\ee
provided
\be\label{49}
g A^2 = \sech^{2}(\beta,m)\,,~~\omega = -2 \cosh(\beta,m)\,,
\ee
while the solution with $B=-A$ goes over to the vacuum solution. 

{\bf Solution V}

Remarkably, unlike the usual focusing AL case \cite{bis}, the same nonlocal, AL 
Eq. (\ref{37}) (which admits $\cn$ and $\dn$ solutions), also admits the periodic 
$\sn$ solution
\be\label{50}
u_{n} = A \sqrt{m}\, \sn(\beta n,m)\, e^{-i(\omega z+\delta)}\,,
\ee
provided
\be\label{51}
g A^2 \ns^{2}(\beta,m)= 1\,,~~\omega = -2\cn(\beta,m) \dn(\beta,m)\,.
\ee

{\bf Solution VI}

In the limit $m=1$, this periodic solution goes over to the hyperbolic kink 
solution
\be\label{52}
u_{n} = A \tanh(\beta n)\, e^{-i(\omega z+\delta)}\,,
\ee
provided
\be\label{53}
g A^2 = \tanh^{2}(\beta)\,,~~\omega = - 2 \sech^{2}(\beta)\,.
\ee

\subsection{Shifted Periodic Solutions}

We now show that this model also admits shifted periodic solutions which we now
discuss one by one. 

{\bf Solution VII}

It is easily checked that
\be\label{54}
u_{n} = \frac{A}{\dn(n \beta,m)}\,e^{-i(\omega z+\delta)}\,,
\ee
is an exact solution to Eq. (\ref{37}) provided
\be\label{55}
g A^2 = \frac{[\cs(\beta,m) \cn^{2}(\beta,m)-2\cs(2\beta,m)\dn(\beta,m)]}
{\cn^{2}(\beta,m)}\,,~~\omega = -\frac{2\dn(\beta,m)}{\cn^{2}(\beta,m)}\,.
\ee

{\bf Solution VIII}

It is easily checked that
\be\label{56}
u_{n} = A \sqrt{m} \frac{\cn(n \beta,m)}{\dn(n \beta,m)}\,e^{-i(\omega z+\delta)}\,,
\ee
is an exact solution to Eq. (\ref{37}) provided
\bea\label{57}
&&\omega \cs^{2}(\beta,m) 
=-2(1+g A^2)\, \ns(\beta,m) [\cs(2\beta,m)+\ds(2\beta,m)]\,,
\nonumber \\
&&\omega \cs(\beta,m)\cs(2\beta,m) \ds(\beta,m) = -\cs^{2}(\beta,m)
[\ds(2\beta,m)+\cs(2\beta,m)] \nonumber \\
&&-g A^2\, \ds^{2}(\beta,m) [
\cs(2\beta,m)+\ds(2\beta,m)]\,.
\eea
On solving these two equations, one can obtain $\omega$ and $gA^2$.

{\bf Solution IX}

It is easily checked that
\be\label{58}
u_{n} = A \sqrt{m} \frac{\sn(n \beta,m)}{\dn(n \beta,m)}\,e^{-i(\omega z+\delta)}\,,
\ee
is an exact solution to Eq. (\ref{16}) provided
\bea\label{59}
&&\omega \cs^{2}(\beta,m) 
=-2(1+g A^2)\, \ds(\beta,m) [\cs(2\beta,m)+\ns(2\beta,m)]\,,
\nonumber \\
&&\omega \cs(\beta,m)\cs(2\beta,m) \ns(\beta,m) = -\cs^{2}(\beta,m)
[\ns(2\beta,m)+\cs(2\beta,m)] \nonumber \\
&&-g A^2\, \ns^{2}(\beta,m) [\cs(2\beta,m)+\ds(2\beta,m)]\,.
\eea
On solving these two equations, one can obtain $\omega$ and $gA^2$.

\section{Coupled Nonlocal Ablowitz Ladik Model}

We now consider a coupled nonlocal AL model which is similar 
but more general than the one considered by us elsewhere \cite{ks6} 
\be\label{7.1}
i\frac{du_{n}}{dz}(z) +[u_{n+1}(z)+u_{n-1}(z)]
+ [a u_{n}(z) u^{*}_{-n}(z)+b v_{n}(z) v^{*}_{-n}(z)][u_{n+1}(z)+u_{n-1}(z)]=0\,,
\ee
\be\label{7.2}
i\frac{dv_{n}}{dz}(z) +[v_{n+1}(z)+v_{n-1}(z)-2v_{n}(z)]
+[d u_{n}(z) u^{*}_{-n}(z)+e v_{n}(z) v^{*}_{-n}(z)][v_{n+1}(z)+v_{n-1}(z)] =0\,.
\ee
It is easy to check that these coupled nonlocal equations will have several
solutions. As an illustration, we now present few such  solutions.

{\bf Solution I}

It is easy to check that 
\be\label{7.3}
u_{n} = A \dn(n \beta,m)\,e^{-i(\omega_{1} z+\delta)}\,,
\ee
\be\label{7.4}
v_{n} = B \sqrt{m} \sn(n \beta,m)\,e^{-i(\omega_{2} z+\delta)}\,,
\ee
is an exact solution to the coupled Eqs. (\ref{7.1}) and (\ref{7.2})
provided Eqs. (\ref{5.5}) are satisfied and further
\be\label{7.6}
\omega_{1} =  -2(1- b B^2) \frac{\dn(\beta,m)}{\cn^{2}(\beta,m)}\,,
~~\omega_{2} = - 2 (1+d A^2)  \cn(\beta,m) \dn(\beta,m)\,.
\ee

{\bf Solution II}

It is easy to check that 
\be\label{7.8}
u_{n} = A \sqrt{m} \cn(n \beta,m)\,e^{-i(\omega_{1} z+\delta)}\,,
\ee
\be\label{7.9}
v_{n} = B \sqrt{m} \sn(n \beta,m)\,e^{-i(\omega_{2} z+\delta)}\,,
\ee
is an exact solution to the coupled Eqs. (\ref{7.1}) and (\ref{7.2})
provided Eqs. (\ref{5.10}) are satisfied and further
\be\label{7.11}
\omega_{1} =  -2 (1-b B^2 m) \frac{\cn(\beta,m)}{\dn^{2}(\beta,m)}\,,
~~\omega_{2} = - 2 (1+d A^2 m) \cn(\beta,m) \dn(\beta,m)\,.
\ee

{\bf Solution III}

It is easy to check that 
\be\label{7.13}
u_{n} = A \dn(n \beta,m)\,e^{-i(\omega_{1} z+\delta)}\,,
\ee
\be\label{7.14}
v_{n} = B \dn(n \beta,m)\,e^{-i(\omega_{2} z+\delta)}\,,
\ee
is an exact solution to the coupled Eqs. (\ref{7.1}) and (\ref{7.2})
provided Eqs. (\ref{5.15}) are satisfied and further
\be\label{7.16}
\omega_{1} = \omega_{2} =  -2 \frac{\dn(\beta,m)}{\cn^{2}(\beta,m)}\,.
\ee

{\bf Solution IV}

It is easy to check that 
\be\label{7.18}
u_{n} = A \sqrt{m} \cn(n \beta,m)\,e^{-i(\omega_{1} z+\delta)}\,,
\ee
\be\label{7.19}
v_{n} = B \sqrt{m} \cn(n \beta,m)\,e^{-i(\omega_{2} z+\delta)}\,,
\ee
is an exact solution to the coupled Eqs. (\ref{7.1}) and (\ref{7.2})
provided Eqs. (\ref{5.20}) are satisfied and further
\be\label{7.21}
\omega_{1} = \omega_{2} =  - 2\frac{\cn(\beta,m)}{\dn^{2}(\beta,m)}\,.
\ee

{\bf Solution V}

It is easy to check that 
\be\label{7.23}
u_{n} = A \dn(n \beta,m)\,e^{-i(\omega_{1} z+\delta)}\,,
\ee
\be\label{7.24}
v_{n} = B \sqrt{m} \cn(n \beta,m)\,e^{-i(\omega_{2} z+\delta)}\,,
\ee
is an exact solution to the coupled Eqs. (\ref{7.1}) and (\ref{7.2})
provided Eqs. (\ref{5.25}) are satisfied and further
\be\label{7.26}
\omega_{1}  =  -2[1-(1-m) bB^2]  \frac{\dn(\beta,m)}{\cn^{2}(\beta,m)}]\,,
~~\omega_{2}  =  -2[1+ (1-m) d A^2]  \frac{\cn(\beta,m)}{\dn^{2}(\beta,m)}]\,,
\ee

{\bf Solution VI}

It is easy to check that 
\be\label{7.27}
u_{n} = A \sqrt{m} \sn(n \beta,m)\,e^{-i(\omega_{1} z+\delta)}\,,
\ee
\be\label{7.28}
v_{n} = B \sqrt{m} \sn(n \beta,m)\,e^{-i(\omega_{2} z+\delta)}\,,
\ee
is an exact solution to the coupled Eqs. (\ref{7.1}) and (\ref{7.2})
provided Eqs. (\ref{5.30}) are satisfied and further
\be\label{7.34}
\omega_{1} = \omega_{2} =  - 2\cn(\beta,m) \dn (\beta,m)\,.
\ee

In the limit $m=1$ these coupled solutions go over to the corresponding hyperbolic 
pulse or kink solutions which can be easily worked out from here.

\section{Conclusion and Open Problems}

In this paper, we have considered a number of discrete, continuum and coupled nonlocal 
nonlinear equations which are non-Hermitian but invariant under PT symmetry. We have 
shown that like the local nonlinear equations (which are Hermitian), these 
nonlocal nonlinear equations 
also admit periodic solutions in terms of $\cn(x,m)$ and $\dn(x,m)$ as well as in  terms
of their linear superposition (and hence the corresponding hyperbolic bright soliton
solution). However, what is remarkable is that unlike the usual local models, all
the six nonlocal models, which admit $\dn(x,m)$ and $\cn(x,m)$ solutions, at the same time also
admit periodic $\sn(x,m)$ and hence hyperbolic dark soliton solution. This
is probably for the first time that the same nonlinear model simultaneously 
admits both the dark and the 
bright soliton solutions. Further, though naively it would seem that no solutions 
with a shift in the transverse coordinate $x$ are allowed in such nonlinear models, 
but in this paper we have shown 
that Jacobi elliptic function solutions with definite shifts are allowed in all the 
six cases. Moreover, in the coupled nonlocal NLSE case, we have 
shown that as in the local case, it admits solutions in terms of Lam\'e polynomials of 
order 1 and 2. What is interesting to note is that while in the local case, for both the 
Manakov and the MZS case,  there are solutions in terms of Lam\'e polynomials of order 
2, in the nonlocal case that we have considered, no solutions in terms of Lam\'e
polynomials of order 2 are admitted in the Manakov case while a large number of 
such solutions are allowed in the MZS case. However, periodic solutions in terms of
shifted Lam\'e polynomials of order 2 do exist in the Manakov case.  

It is worth pointing out here that in this paper we have restricted ourselves to
only the nonsingular solutions. In fact several singular solutions are admitted as well 
in all the six cases [like $\cn(x,m)/\sn(x,m)$) but we do not consider them here. 

Finally, it is worth enquiring whether, like the nonlocal NLSE, the coupled nonlocal NLSE
of Manakov type or MZS type are also integrable systems? Looking at the structures
of the nonlocal NLSE and the coupled nonlocal NLSE, we conjecture that indeed both 
nonlocal Manakov and nonlocal MZS are also likely to be integrable systems. One reason 
why we think so is because the conserved quantities in both the uncoupled and the coupled 
nonlocal NLSE models are simply obtained from the corresponding local case by 
replacing $u^{*}(x,z)$ by $u^{*}(-x,z)$. Remarkably, a similar situation also occurs
for the nonlocal Ablowitz-Ladik case and hence we conjecture that the nonlocal 
Ablowitz-Ladik system is also likely to be integrable. It will be interesting to 
prove or disprove our conjectures. Further, it would be worthwhile to study the 
stability of the various soliton solutions of the system.

\section{Acknowledgments} This work was supported in part by the U.S. Department 
of Energy.

\end{document}